\documentclass[12pt, a4paper]{article}
\usepackage[utf8]{inputenc}
\usepackage{dcolumn,lscape}
\usepackage{amsmath,longtable,multicol,dcolumn,tabularx,amssymb}
\usepackage{exscale,amsthm,multirow,rotating}
\usepackage{natbib}
\usepackage{bbm}
\usepackage{tikz,dsfont,float}

\usepackage{todonotes}

\usepackage{hyperref}

\hypersetup{
    colorlinks=true,
    linkcolor=blue,
    citecolor=blue,
    filecolor=magenta,      
    urlcolor=cyan,
    pdftitle={review paper},
    pdfpagemode=FullScreen,
}

\newcommand{\xvec}{\boldsymbol}
\newcommand{\xmat}{\mathbf}
\newcommand{\xset}{\mathds}

\newcommand{\blue}{\textcolor{blue}}		

\usepackage{graphicx}
\usepackage{caption}
\usepackage{subcaption}

\usepackage{changes}
\definechangesauthor[name={PM}, color=blue]{PM}
\definechangesauthor[name={MP}, color=red]{PO}
\definechangesauthor[name={AF}, color=violet]{AF}

\graphicspath{{"Figures/"}}

\begin{document}

\def\spacingset#1{\renewcommand{\baselinestretch}%
	{#1}\small\normalsize} \spacingset{1}


\title{\bf A review of regularised estimation methods and cross-validation in spatiotemporal statistics}
\author{Philipp Otto\footnote{Corresponding author: philipp.otto@glasgow.ac.uk}\\
	\small{University of Glasgow, School of Mathematics and Statistics, United Kingdom}\\
	Alessandro Fass\`{o}\\
	\small{University of Bergamo, Italy}\\
	Paolo Maranzano\\
	\small{Department of Economics, Management and Statistics (DEMS), University of Milano-Bicocca, Italy} \\ \& \small{Fondazione Eni Enrico Mattei (FEEM), Italy}}
\maketitle

\begin{abstract}
	This review article focuses on regularised estimation procedures applicable to geostatistical and spatial econometric models. These methods are particularly relevant in the case of big geospatial data for dimensionality reduction or model selection. To structure the review, we initially consider the most general case of multivariate spatiotemporal processes (i.e., $g > 1$ dimensions of the spatial domain, a one-dimensional temporal domain, and $q \geq 1$ random variables). Then, the idea of regularised/penalised estimation procedures and different choices of shrinkage targets are discussed. Finally, guided by the elements of a mixed-effects model setup, which allows for a variety of spatiotemporal models, we show different regularisation procedures and how they can be used for the analysis of geo-referenced data, e.g. for selection of relevant regressors, dimensionality reduction of the covariance matrices, detection of conditionally independent locations, or the estimation of a full spatial interaction matrix.
\end{abstract}

\noindent%

\spacingset{1.45} 

\newpage

\section{Introduction}

In the era of big geospatial data, the analysis of intricate spatial and spatiotemporal processes has become both increasingly vital and challenging. This is attributed to the growing volume and finer resolution of geo-referenced data, including remotely sensed, crowd-sourced or LiDAR data, as well as the increasing diversity of data types due to automated collection and processing, such as social network data and image data. These advancements enable the examination of more intricate relationships and allow for the detection of weaker dependencies or variations in the data. To statistically investigate such interactions, geostatistical and spatial econometric models offer a robust framework for comprehending the fundamental structures of these processes. Additionally, they provide insights into spatial dependencies, temporal dynamics, and the interactions among multiple random variables. An essential advantage of these statistical models lies in their interpretability, unlike deep learning models, which are often viewed as black-box models unless explicitly designed for interpretability. However, as the dimensionality of the spatial domain increases and datasets become larger and more complex, conventional modelling approaches frequently encounter challenges related to computational complexity and the appropriate selection of influential variables.

{In this review paper, our focus is on regularised estimation procedures for spatiotemporal continuous data, mainly focusing on Gaussian geostatistical models and spatial autoregression models{.The latter ones are often called spatial econometrics models, but we will refer to them as spatial autoregression models because 
the response variable is explicitly correlated with its adjacent observations in an autoregressive manner, and} they are not only tied to applications in economics. Thus, these models are typically used for data 
{defined} on a discrete set of spatial locations, e.g., regular or irregular lattices, including polygons such as municipalities, counties or countries, making them attractive modelling approaches in econometrics (see, e.g., \citealt{bille2019spatial,anselin2010thirty,arbia2016spatial}), ecology (see, e.g., \citealt{ver2018spatial,lichstein2002spatial}), or epidemiology (see, e.g., \citealt{lee2011comparison,gebreab2018statistical}). 
{Instead}, in geostatistics, spatial dependence is modelled using covariance functions depending on the distance between two observations\footnote{Note that the distance is not necessarily measured as Euclidean distance, but different distance measures such as great-circle distances for modelling processes on a sphere or suitable network distances for processes on networks can be considered.}. This makes the model attractive for {processes on a continuous space} and 
data which is irregularly observed across space, such as air pollution at ground-level measurement stations. Our focus will be mainly on stationary processes (across space and time) at a fixed set of spatial locations. Spatial point processes, such as Poisson processes, will not be the focus of this review paper (see \citealt{gonzalez2016spatio} for a review on spatial point processes).}



Regularised estimation procedures offer effective solutions for these challenges from multiple perspectives. For instance, they can be used for model selection and dimensionality reduction but also to reveal spatial dependence structures going beyond geographical proximity. This article will provide a comprehensive review of regularisation techniques, including shrinkage and penalisation methods, used for statistical modelling of geospatial data. Thereby, our focus will be on geostatistical models (Section \ref{sec:geostat}) and spatial \blue{autoregression} models (Section \ref{sec:spatecon}). Since the degree of regularisation is typically chosen based on predictive accuracy, we also review cross-validation techniques which are suitable under spatiotemporal dependence (Section \ref{sec:cv_strategies}). Finally, Section \ref{sec:conclusion} concludes the article and discusses future research directions.

\section{General framework}

Assume that we are observing a continuous $q$-dimensional vector over space and time. The related $q$-variate dataset may be formalised as 
\begin{equation}
    \{y_{t_j}(\xvec{s_i}) \in \xset{R}^q: i=1,...,N, j =1,...,T\}
    \label{eq:dataset}
\end{equation}
where $\xvec{s_i}$ are points in the spatial domain $D_{\xvec{s}}$, which may be a Euclidean or non-Euclidean space. Common choices are the plain $ \xset{R}^2$, the Earth sphere $\xset{S}^2$ or a discrete grid of points.
The time domain, say $D_t \ni t_i$, is assumed discrete and, ignoring missing values, is made by equidistant time points. For simplicity, we use the set of integers,
$D_t=\xset{Z}$ and $t_j=j=1,...,T$.
In the statistical framework, data $y_t(s_i)$ are assumed to be generated by the spatiotemporal stochastic process
$\{Y_t(\xvec{s}) \in \xset{R}^q: \xvec{s} \in D_{\xvec{s}}, t \in D_t \}$, \citep[see][]{cressie2015statistics}. A continuous temporal domain is sometimes considered in geostatistics, see, e.g., \cite{PorcuWhite2022}, but we focus on the first case, which may be called the spatial time series approach.

\begin{figure}
	\centering
	\includegraphics[width=14cm ,height=10cm]{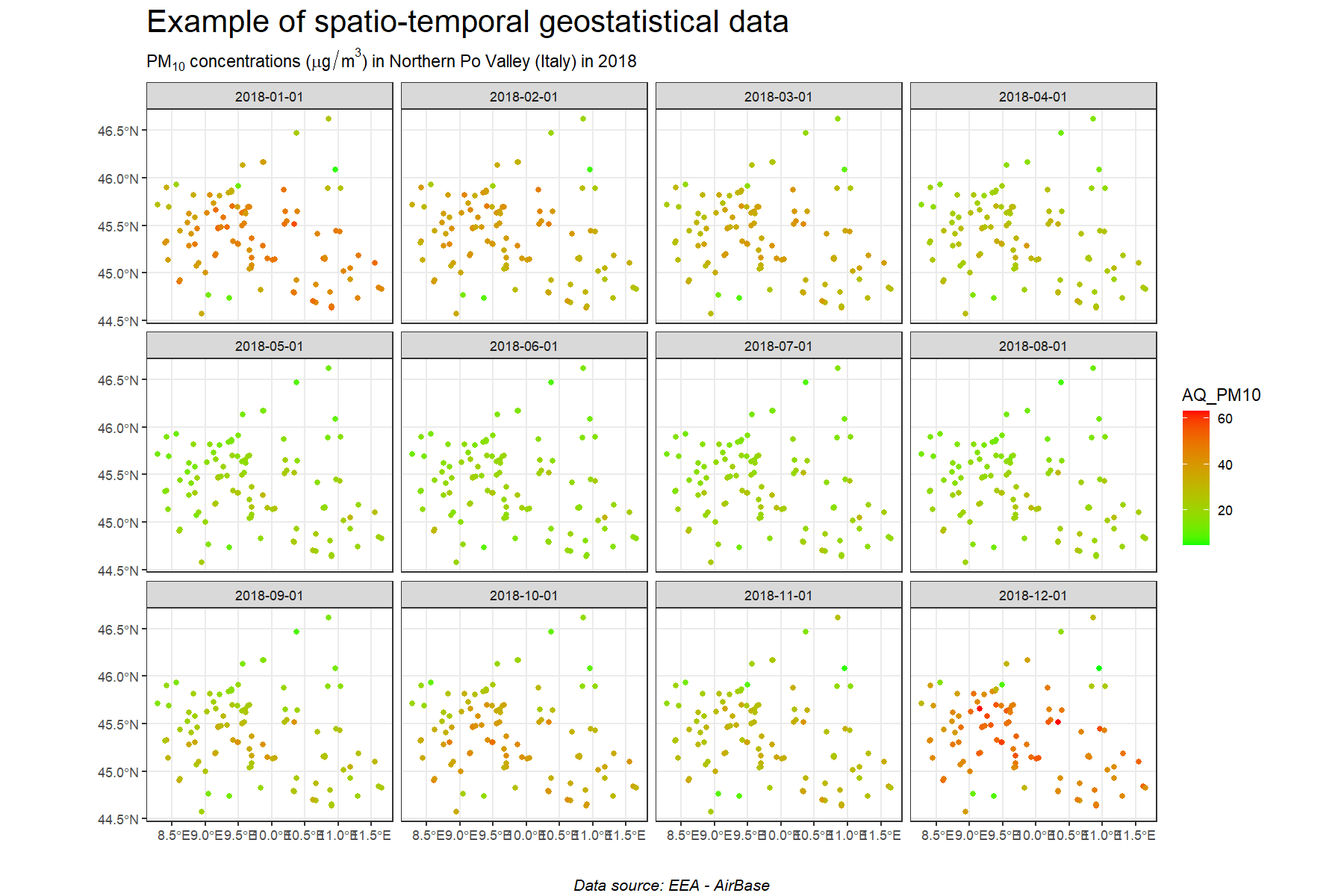}
	\caption{Illustrative example of real-world geostatistical spatiotemporal data. The plot shows the observed PM$_{10}$ concentrations in 2018 at the 101 locations in Northern Italy considered by \cite{Agrimonia2023}. For each panel, the observed average monthly concentrations are reported.}
	\label{fig:Ex_STgeostat}
\end{figure}

\begin{figure}
	\centering
	\includegraphics[width=14cm ,height=10cm]{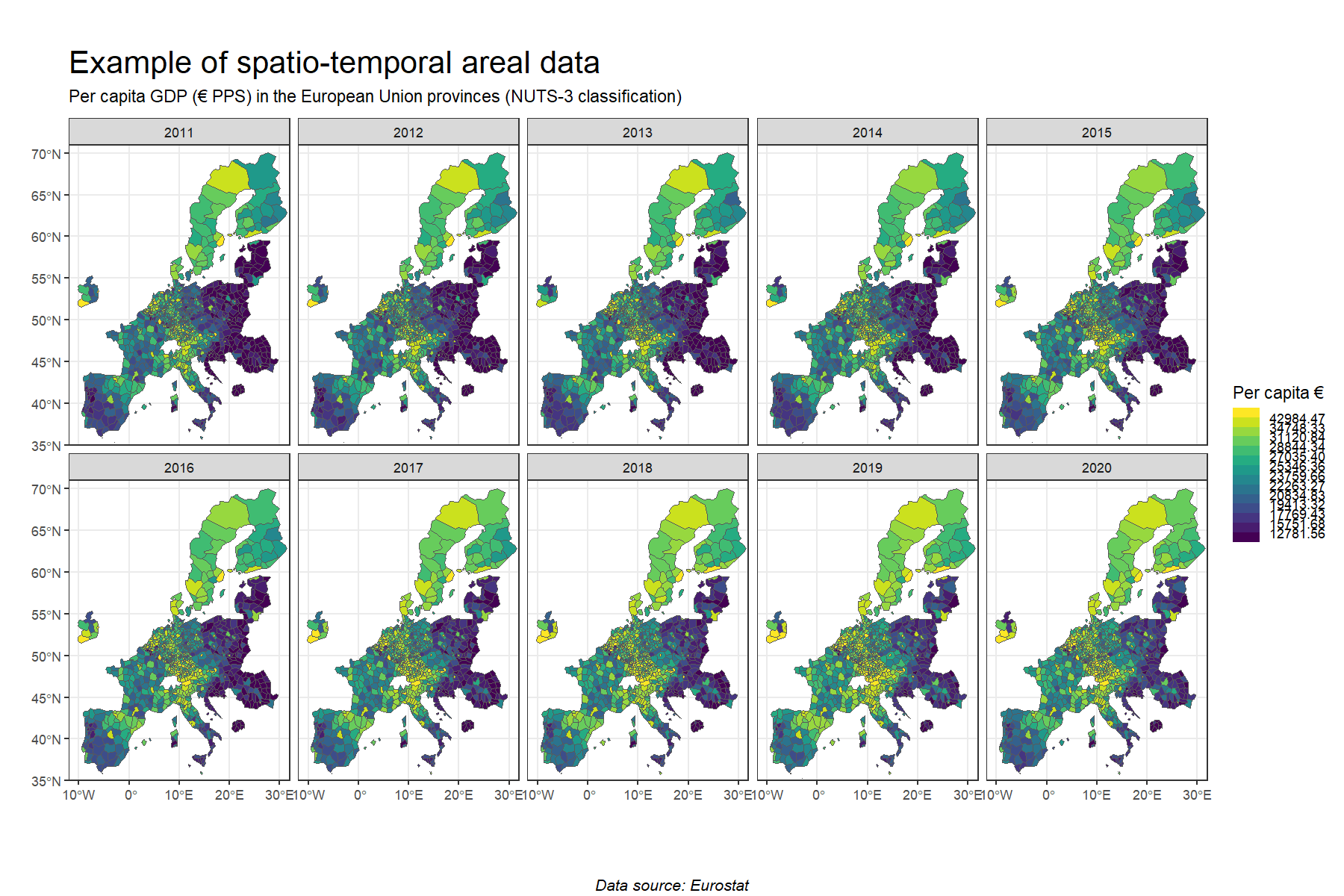}
	\caption{Illustrative example of spatiotemporal areal (irregular lattice) data. The plot shows the estimated yearly per capita gross domestic product (€ per capita) in the 1088 European Union provinces (NUTS-3 classification) from 2011 to 2020. Data source: Eurostat}
	\label{fig:Ex_STareal}
\end{figure}

The above-mentioned spatial and temporal domains can assume several forms, resulting in multiple data categories. In particular, following \cite[Section 2.1]{ZammitWikleCressie2019}, we can distinguish among three main classes of spatiotemporal data. By geostatistical data, we mean phenomena which can be measured at continuous locations over a given spatial domain. Typical examples are air quality, weather and climate measurements, such as the PM$_{10}$ concentrations reported in Figure \ref{fig:Ex_STgeostat}. By areal or lattice data, we mean a phenomenon defined on a finite or countable subset in space over a specific time span. Examples of areal data are easily found in the socio-economic, political or medical-epidemiological realms, where observations are often reported by area (e.g., municipalities or regions). For example, in Figure \ref{fig:Ex_STareal}, we show the evolution of per capita income for the European provinces (NUTS-3 classification) between 2011 and 2020. They provide a common playground for statisticians, econometricians, or applied social scientists, as the neighbourhood structure among units can provide relevant insights in many empirical contexts while preserving the interpretability of the phenomenon. 

There are different sources of correlation for such geo-referenced data:
\begin{enumerate}
\item 	Temporal proximity between two observations typically causes correlations. That is, the closer two observations are in the temporal domain $D_t$, the higher they are correlated, e.g., the current outside temperature will be similar to the temperature one hour earlier but less similar to the temperature two days ago.

Also, more complex time-related structures are possible. For example, considering periodicity, the daily and seasonal cycles often require non-monotonic correlation structures.
\item Geographical proximity in $D_{\xvec{s}}$ induces a similar dependence, known as spatial dependence. As claimed by Tobler's first law of Geography, two observations are more similar if they are close to each other. For example, the current outside temperature in a city is similar to that in the neighbouring cities but less similar if you move away, especially in the North-South direction. 
\item The $q$ variables could be cross-correlated. Otherwise, if the $q$ variables are (statistically) independent, one would rather prefer to model them independently using $q$ univariate models. Moreover, network structures could describe the interrelations between these variables. In this case, network or graphical models for spatiotemporal data are suitable \citep[see, e.g.,][for graphical Gaussian processes for multivariate spatial data]{dey2020graphical,zapata2022partial}. However, graphical models are beyond the scope of this paper.
\end{enumerate}

There are two alternative, complementary approaches in spatial and spatiotemporal statistics to account for spatial dependence. Firstly, the dependence can be modelled using spatially dependent processes, where a suitable covariance function defines the entries of the covariance matrix. This approach is commonly known as geostatistics. Secondly, the dependent/outcome variable can be explicitly correlated with the nearby observations. Thereby, the local neighbourhood is defined by a suitable weight matrix $\xmat{W}$, which is weighting all other observations, such that the product of $\xmat{W}$ and the dependent variable is the weighted average of the adjacent observations. Below, we will provide an overview of regularised estimation procedures for both approaches, starting with geostatistical models in Section \ref{sec:geostat} and followed by spatial autoregression in Section \ref{sec:spatecon}.

\section{Geostatistical models}\label{sec:geostat}


A quite general frame for modelling data in \eqref{eq:dataset} is given by the mixed-effects spatiotemporal model, which is given by
\begin{equation}
	Y_t(\xvec{s}) = \mu_t(\xvec{s}) + \omega_t(\xvec{s}) + \varepsilon_t(\xvec{s}) ,
\end{equation}
where $\mu_t(\xvec{s})$ is the fixed-effects component, $\omega_t(\xvec{s})$ is the random-effects component, and $\varepsilon_t(\xvec{s})$ is the stochastic error at time $t$ and location $\xvec{s}$, which are typically assumed to follow a (Gaussian) white noise process. While the fixed-effects term typically models the mean behaviour and the influence of exogenous variables, the random-effects term accounts for any additional dependence in the data, including temporal and spatial dependence, heterogeneity, and cross-correlation between the variables. {We will follow this distinction throughout the paper and discuss regularised estimation procedures for the parameters of $\mu_t(\xvec{s})$ and $\omega_t(\xvec{s})$. Penalised estimation procedures can be used to shrink certain model parameters towards a pre-specified target. The most often applied shrinkage target is zero. In this case, the parameters are shrunk towards zero, which means that a parameter is excluded from the model if its estimate is equal to the zero target. Thus, these methods are suitable for model selection (simultaneous parameter estimation and model selection). In particular, when applied to the fixed-effects part, they can automatically select relevant regressors. Compared to standard model selection procedures, such as step-wise selection based on cross-validation goodness-of-fit measures, regularised estimation procedures are typically computationally more efficient\footnote{Note that only one or a few hyper-parameters, i.e., regularisation parameters, need to be selected via cross-validation, while for other complete or stepwise selection procedures, significantly more cross-validation iterations are needed.}.}

Generally, there are two main sources of increasing computational complexity: model dimension and data dimension.
The former is related to the fixed-effects component when considering variable selection. This is developed in Section \ref{sec:regularisation_largescale}. Moreover, the model dimension is related to the random-effects component when modelling spatial and temporal correlations, which are considered in Section \ref{sec:regularisation_geospat}.

Considering data dimension, the observation covariance matrix of the data set (\ref{eq:dataset}) is $qNT\times qNT$ dimensional, and its brute-force inversion has a cubic computational cost. Hence, specialised modelling and computing techniques are needed. The leading term in modern large spatial data sets is the number of spatial locations $N$. Fortunately, the correlation often decreases with an increasing distance between observations, allowing for a sparse covariance or precision matrix representation. {Regularised estimation procedures with a shrinkage target of zero can be used to introduce zeros in the spatial covariance matrix or the precision matrix, indicating conditional independence between these observations. This will be the subject of Section \ref{sec:regularisation_geospat}.}
Similarly, if the number of variables $q$ increases, we observe a corresponding cubic increase characterised by the fact that the cross-correlation matrices for contemporaneous and colocated observations are usually dense.
On the other side, the temporal dimension is often less critical because causal time series models may leverage on the fact that only past observations influence future observations, and the computational complexity may be reduced to be linear in $T$, see e.g. \cite{DSTEMv2}, §2.4. These issues are {also} considered in Section \ref{sec:regularisation_geospat}.


\subsection{Regularisation for the fixed-effects component}\label{sec:regularisation_largescale}

Linear regression models, whether referring to spatiotemporal or unstructured data, can be estimated using various techniques, including Likelihood maximisation and least squares. However, in high-dimensional settings, traditional methods for regression models might not be directly applied \citep{NandyEtAl2017}. In order to reduce the model's complexity due to the fixed effect component $\mu_t(\xvec{s})$, several methods for spatial and spatiotemporal settings have been proposed. Particularly, we refer to the task of selecting the most relevant predictors among a large set of candidates.

The state-of-the-art statistical literature inherent to variable selection pivots around the Least Absolute Shrinkage and Selection Operator (LASSO) approach introduced by \cite{tibshirani1996regression}. LASSO is a penalised version of ordinary least squares that uses a LASSO-type $l_1$ penalisation to shrink irrelevant parameters to zero. Under mild regularity conditions, including uncorrelated observations, LASSO ensures consistent parameter estimation \citep[see Section 2.4.2 of][]{BuhlmannVanDeGeer2011} and model selection consistency \citep{BickelEtAl2009,BelloniChernozhukov2013}, i.e., LASSO owns the oracle property. Among others, one of the most relevant features of LASSO is that it admits solutions to the minimisation task even when the number of parameters is greater than the number of available observations. Also, LASSO penalty (and extensions) can be used either considering least squares and penalised likelihood problems \citep{FanLi2001}.

In the field of spatial and spatiotemporal regression, the number of observations can be very large, and the likelihood computation, even for Gaussian models, can require a very high computational effort \citep{SteinEtAl2004,Stein2014}. Thus, when it comes to maximum likelihood estimation of space-time models, a major pathway is to rely on penalised maximum likelihood estimators (PMLE) of the parameters aiming at approximating the true likelihood function \citep{FanLi2001,FanPeng2004,ZouLi2008}. However, asymptotic properties of approximate PMLEs rely on the asymptotic distribution of the initial estimators (e.g., ML estimates) used in the optimisation algorithm \citep{Liu2017}. To further improve computational efficiency, \cite{ChuZhuWang2011} proposed a geostatistical version of PMLE in which the penalised function is approximated using one-step sparse estimator \citep{ZouLi2008} and covariance tapering \citep{FurrerEtAl2006}.

It is worth noting that real-world geostatistical applications are prone to cross-correlated regressors due to their dependent (spatially and temporally) structure. \cite{zhao2006model} point out that when predictors are correlated, the classic LASSO algorithm does not provide selection-consistent estimates. Furthermore, when cross-correlation is detected, also group-LASSO estimators, which assume orthonormal data within each group, perform poorly in selecting the relevant predictors \citep{SimonTibshirani2012}. A straightforward solution to this issue is provided by the adaptive LASSO penalty, which leads to selection-consistent estimators even in the presence of cross-correlated covariates \citep[see][]{zou2006adaptive} and \cite{ZouLi2008}. 

In addition to penalised likelihood methods, other estimation techniques have been developed for LASSO and its extensions in the geostatistical domain. For example, penalised least squares algorithms have been extended to the case of linear spatial models \citep{WangZhu2009}, spatial autoregressive models \citep{CaiMaiti2020}, regression models with spatially dependent data \citep{HuangEtAl2010} and conditional autoregressive models \citep{GonellaEtAl2022}. Also, when considering additive spatial models with potential non-linear effects, weighted versions of penalised least squares can be applied \citep{NandyEtAl2017}. \cite{CaiEtAl2019} proposed a generalised method-of-moments LASSO, which combines LASSO with GMM estimator, to perform variable selection for spatial error models with spatially autoregressive errors. \cite{HuangEtAl2010} propose using LASSO to simultaneously select relevant predictors, choose neighbourhoods, and estimate parameters for spatial regression with GIS layers to predict responses in unsampled sites. \cite{SafikhaniEtAl2020} considered LASSO methods for generalised spatiotemporal autoregressive models. The estimators are obtained by a modified version of the penalised least squares that accommodates hierarchical group LASSO-type penalties.\cite{ChernozhukovEtAl2021} combine least squares LASSO and bootstrap procedures to get estimates and inference for systems of high-dimensional regression equations characterised by temporal and cross-sectional dependences in covariates and error processes. \cite{CaoEtAl2022} proposed a penalised estimation procedure for Gaussian Processes regressions where the likelihood and the first two derivatives are approximated by means of a scaled Vecchia approximation \citep{vecchia1988estimation}. Eventually, several application-oriented papers combine classic LASSO approaches and geostatistical models in multi-step procedures \citep[e.g.,][]{SPASTA2022,ye2011sparse,pejovic2018sparse}.

Penalised methods are also commonly applied in the context of functional data analysis, especially involving penalised splines \citep[see][]{RamsaySilverman2002}. These methods usually regularise the smoothness of the estimated functions by penalising the integrated second derivatives. In this way, many basis functions can be used, thus avoiding the typical overfit resulting from unpenalised estimation methods. Several authors have attempted to contribute by proposing LASSO-like penalised methods for selecting relevant functional (group) predictors \citep{PannuBillor2017} or to identify regions where the coefficient function is zero and to smoothly estimate non-zero values of the coefficient function \citep{CentofantiEtAl2022}. Basis expansions and low-rank representations \citep{Wood2017} are widely used tools in geostatistics for the spatiotemporal interpolation of environmental phenomena  \citep[see][for group-LASSO approaches in this context]{HofierkaEtAl2002,XiaoEtAl2016,ChangHsuHuang2010}. For instance, \cite{SERRA2023} use a PMLE with an adaptive LASSO penalty to select relevant functional covariates or their statistically relevant regions using the hidden dynamic geostatistical model. Eventually, \cite{HsuChangHuang2012} deal with semiparametric models for non-stationary spatiotemporal data in which a penalised least square with group-LASSO penalty is used to identify local spatial-temporal dependence features deviated from the main stationary structure.

\subsection{Regularisation for the random-effects component}\label{sec:regularisation_geospat}

Suppose that the random-effects component follow a {stationary} $q$-variate Gaussian process, i.e.,
\begin{equation}
	\{\omega_t(\xvec{s}) : \xvec{s} \in D_{\xvec{s}}, t \in D_t \} \quad \sim \quad N_p(0, C_\theta(\xvec{s} - \xvec{s}', t - t')) ,
 \label{eq:rand.eff}
\end{equation}
where $C_\theta$ is a matrix covariance function that depends on the difference between any two locations $\xvec{s}$ and $\xvec{s}'$ and two arbitrary time points $t$ and $t'$. It is the fundamental building block of the covariance matrix of the data set (\ref{eq:dataset}) and, for $q > 1$, also provides the cross-covariances between the components of the observation vector $y$. In order to have a positive semidefinite observation covariance matrix, $C_\theta$ must be a ``valid'' covariance function. See \citealt{gneiting2002nonseparable, gneiting2010matern, stein2005space, nychka2002multiresolution, porcu2016spatio} or \citealt{porcu202130} for a historical review of space-time covariance functions.  The covariance function in (\ref{eq:rand.eff}) usually depends on some unknown parameters $\theta$, which have to be estimated. 


Generally, it is desirable if the covariance matrix resulting from $C_\theta$ contains many zeros; that is, it is sparse. The traditional way to induce zeros in the covariance is known as covariance tapering \citep{furrer2006covariance}. Loosely speaking, based on the geographical distance between the locations, a covariance of zero is assumed for observations whose distance is larger than a certain threshold. Theoretical results on covariance tapering can be found in \citealt{stein2013statistical}, and the use of tapering in the multivariate case is considered in \cite{Bevilacqua2016}. This method can also be applied in likelihood-based estimation procedures (see \citealt{kaufman2008covariance}, and \citealt{furrer2016asymptotic} for theoretical results). As mentioned above, regularised estimation procedures are tailor-made to induce zeros for certain parameters when the shrinkage target is chosen to be zero. Typically, we can find zeros if two observations across space/time are conditionally independent (i.e., independent when observing all other realisations). However, the conditional independence is encoded in the inverse covariance matrix or precision matrix. Thus, penalised methods can be applied to obtain sparse precision matrices. For univariate data ($q = 1$), \cite{krock2021nonstationary} introduced a graphical LASSO procedure to induce zeros in the precision matrix of a spatial process. \cite{krock2021modeling} extended this approach to be used in the multivariate case ($q > 1$). 

Moreover, since the covariance matrix is a positive definite matrix by definition, both the covariance matrix and its inverse can be decomposed as $\xmat{\Sigma} = \xmat{P}\xmat{P}'$, e.g., via Cholesky decomposition. The matrix $\xmat{P}$ is called Cholesky factor. \cite{stein2004approximating} proposed to approximate the likelihood for large spatial data sets based on Vecchia approximations \citep{vecchia1988estimation}. The idea is to approximate the joint likelihood as a product of the conditional likelihoods. Now, sparse Cholesky factors have to be considered first by \cite{schaefer2021sparse} for spatial models. If the dimension of $\xmat{P}$ is $N \times r$ with $N$ being the total number of observations across space/time and $r \ll N$, the covariance matrix is low-rank. That is, the dependence structure reduces to a lower-dimensional space. In spatial and spatiotemporal statistics, low-rank covariance matrices have been considered first by \cite{banerjee2008gaussian}, \cite{cressie2008fixed} (spatial fixed-rank kriging), and \cite{cressie2010fixed} (spatiotemporal fixed-rank filtering). Back to the subject of this review paper, \cite{ChangHsuHuang2010} proposed a penalised estimation procedure to identify the lower rank of the covariance matrix. Specifically, low-rank approximations aim to represent a spatiotemporal process as a linear combination of local basis functions, which are weighted by uncorrelated random-effect coefficients (see also the recent review by \citealt{cressie2022basis}). Thus, regularised estimation procedures aim to select suitable local basis functions. From a practical perspective, many different local basis functions (e.g., on several grids with different resolutions) can be included, and the best basis functions are chosen automatically by estimating the model parameters. For the spatiotemporal case, \cite{HsuChangHuang2012} suggested penalised procedures to choose these local basis functions. Furthermore, \cite{kang2021correlation} considered sparse inverse Cholesky factors that are identified based on correlations.

\subsection{Bayesian estimation procedures}

Along with the frequentist paradigm, the literature pioneered penalised regression extensions following a Bayesian perspective. Bayesian estimation schemes have gained particular importance for spatiotemporal models, especially using integrated nested Laplace approximations (INLA), which makes them applicable also for large data sets \citep[see][for a review on INLA in the spatiotemporal context]{rue2017bayesian}. Moreover, we refer the readers to the review paper by \cite{vanErpEtAl2019} for a comprehensive overview of the state-of-the-art literature on Bayesian penalised regression. The authors summarise that Bayesian penalisation techniques include the penalised in three alternative ways. The first way is called \textit{full Bayesian} or \textit{hierarchical Bayesian} \citep[][]{WolpertStrauss1996} approach, which treats the penalty parameter $\lambda$ as an unknown variable (i.e., a hyperparameter) whose prior distribution has to be specified. Such models specify prior distributions for all parameters and can be estimated in a single step. The prior distribution is called the shrinkage prior \citep{vanErpEtAl2019} and is usually a vague distribution, e.g., a half-Cauchy random variable \citep{PolsonScott2012}. The prior acts on the coefficients in order to shrink small effects to zero while maintaining true large effects. Indeed, large values of $\lambda$ result in smaller prior variation and thus more shrinkage of the coefficients towards zero.

The second way is named \textit{empirical Bayesian} approach \citep{vandeWielEtAl2019} and intends the parameters as unknown constants. This approach differs from the first one because it involves a two-step process: first, estimating the penalty parameter $\lambda$ from the observed data, and second, incorporating this empirical estimate into the model using an empirical Bayes prior distribution. Since the empirical approach does not specify any prior distribution for the hyperparameters, a sensitivity analysis of the results with respect to the distributions is not necessary.

The third approach is based on cross-validation (CV). In this case, there is no difference between the frequentist and Bayesian frameworks, as the goal is to select $\lambda$ so that the model is as accurate as possible in predicting new values of the response variable. The topic of spatiotemporal CV will be extensively discussed in the following Section \ref{sec:cv_strategies}.

As mentioned above, the fully Bayesian approach requires defining an a priori distribution for the penalty term of each hyperparameter. Different distributions were proposed depending on the penalised regression type (e.g., LASSO, ridge, or elastic net). For example, in the case of ridge regression \citep{Hastie2020}, the ridge prior corresponds to a Normal centred on the origin \citep{Hsiang1975}, while in the case of LASSO, a Laplace distribution is employed \citep{ParkCasella2008}. For a Bayesian LASSO Gibbs sampler, the Laplace distribution can be represented as a scale mixture of Gaussians (with an exponential mixing density). It is worth noting that the ridge regression estimator can be viewed as the Bayesian posterior mean estimator of the coefficients when imposing a Gaussian prior on the regression parameter \citep{vanWieringen2015}. Further extensions can be found in \cite{vanErpEtAl2019}, in which the authors compare several shrinkage priors from theoretical and application perspectives.

In addition to the penalised regression approach discussed above, the Bayesian framework includes other model selection techniques, such as the Zellner's \textit{g}-prior \citep{Zellner1986}, and the \textit{spike-and-slab prior} \citep{MitchellBeauchamp1988}. The former approach shrinks the regression coefficients toward zero through a global shrinkage scalar called \textit{g}, which equally shrinks each coefficient (which can be reasonable if the coefficients are equivalent). The single-\textit{g} Zellner's prior has been extended in several ways, e.g. using a mixture of \textit{g} priors \citep{LiangEtAl2008}, and multiple shrinkage factors as in \cite{ZhangEtAl2016}. However, carefully choosing the constant \textit{g} must be addressed to avoid excluding important variables \citep{Lindley1957}. Instead, according to the data, the spike-and-slab method assigns the regression coefficients to the zero-centred spike (i.e., shrinking toward zero) if they do not deviate substantially from zero. In contrast, if they differ significantly from zero, they will be assigned to the slab (i.e., the vague proper prior). Both methods were also adapted to spatiotemporal analysis. Refer, for example, to \cite{LeeEtAl2014} on the combination of Zellner's \textit{g} prior and spatial Ising prior for selecting spatial covariates in spatial time series data.

The literature on Bayesian modelling for spatiotemporal data addresses the analysis through both a fully Bayesian hierarchical approach \citep{WikleEtAl1998} and an empirical Bayes approach \citep{FahrmeirEtAl2004}. Fully Bayesian frameworks were extended to the case of variable selection in large spatiotemporal models in several ways. \cite{KatzfussCressie2012} proposed a Bayesian hierarchical spatiotemporal random effects where dimensionality reduction is achieved by means of spatiotemporal basis functions, whereas the prior induces sparsity and shrinkage on the first-order autoregressive parameters describing the temporal evolution of the basis-function coefficients. The described approach was inspired by the so-called \textit{Minnesota prior} \citep{IngramWhiteman1994,GeorgeEtAl2008}. Such prior was initially developed in a time series context where the aim was to drop the autoregressive coefficients of VAR models by shrinking the posterior of the parameter matrix towards independent random walk models (the typical behaviour of stock prices in financial applications).

Possible alternatives to penalty methods for selecting linear predictors in space-time models are mixture model selection methods. This category includes Bayesian selection methods and Bayesian model averaging. The former evaluates the appropriateness of a model based on the estimated weight among a variety of models with alternative predictors; the latter, on the other hand, averages over several alternative models to find the posterior distribution of the parameters. Spatial dynamics is included in the algorithms by adding an intrinsic conditional autoregressive (ICAR) \citep{BYM1991,BesagGreen1993} approach \citep{CarrollEtAl2018}, while temporal dynamics is incorporated via autoregressive processes \citep{LawsonEtAl2017}. Both methods have proven effective in disease mapping studies with spatial data \citep{CarrollEtAl2018}, spatial small area frameworks \citep{CarrollEtAl2016_AoE}, as well as spatiotemporal disease mapping \citep{CarrollEtAl2016_Env}. Among others, relevant advantages include getting an automatic final model fit.

\section{Spatial autoregression}\label{sec:spatecon}

Since this approach requires the explicit definition of the neighbourhood structure via $\xmat{W}$, the prediction at unknown locations is more complicated (it would require a distance-dependent functional relation of each weight), and the models are usually applied to panel data. That is, the observational sites are typically constant across time. Suppose that there are $N$ different locations $\xvec{s}_1, \ldots, \xvec{s}_N$ and $\xvec{Y}_t = (Y_t(\xvec{s}_1), \ldots, Y_t(\xvec{s}_N))'$, then a simple spatial autoregressive process (or simultaneous autoregressive process) without a temporal autoregressive dependence would be given by
\begin{equation}
	\xvec{Y}_t = \xmat{X}_t \xvec{\beta} + \rho \xmat{W} \xvec{Y}_t  + \xvec{\varepsilon}_t \qquad \text{for $t = 1, \ldots, T$}\, ,
\end{equation}
where $\xmat{X}_t$ is a matrix of regressors, $\xvec{\beta}$ is the corresponding vector of regression coefficients, $\rho$ is the spatial autoregressive parameter, and $\xvec{\varepsilon}_t$ is the vector of \blue{white noise} model errors. 

\subsection{Regularisation for the mean model}

For spatial autoregressive (SAR) and conditional autoregressive (CAR) models, \cite{GonellaEtAl2022} proposed a LASSO estimation procedure to select the relevant covariates in the regression term. Among others, \cite{WenEtAl2018} use a spatial autoregressive model with adaptive LASSO penalty to detect relevant autoregressive parameters in genetic studies. \cite{ZhuHuangReyes2010} implemented an iterative penalised likelihood estimator with adaptive LASSO penalty to select predictors and neighbourhood structure in conditional autoregressive and simultaneous autoregressive models with spatially correlated error terms. Later on, \cite{ReyesEtAl2012} proposed an adaptive LASSO algorithm for the case of linear regression models with spatiotemporal neighbourhood structures. \cite{Liu2017} extended the previous algorithms, allowing for spatial correlation to be captured by either the spatial lag terms or spatial errors or both through a SARAR model. Also, their penalised estimates are obtained via least squares approximation to account for possible non-concavity of the likelihood function. Other examples of penalised likelihood for spatiotemporal data are in \cite{AlSulamiEtAl2019}, in which an adaptive LASSO method is proposed to simultaneously identify and estimate spatiotemporal lag interactions in the context of a data-driven semiparametric nonlinear model. Similarly, \cite{Liu2022} developed an adaptive LASSO variable selection method for semiparametric spatial autoregressive panel models with random effects. The estimation is performed by maximising the concentrated profile likelihood function by means of a non-linear optimisation algorithm. Eventually, \cite{ChangHsuHuang2010} and \cite{HsuChangHuang2012} additionally reduced the model's complexity by combining covariance tapering and PMLE for spatial and spatiotemporal settings, respectively.

\subsection{Regularisation for the spatial dependence structure}

Assuming normal random errors with covariance matrix $\xmat{\Sigma}_{\varepsilon}$, this approach can also be considered as a mixed-effects spatiotemporal model with
\begin{eqnarray*}
	\xvec{\mu}_t = (\mu_t(\xvec{s}_1), \ldots, \mu_t(\xvec{s}_n))' & = & (\xmat{I} - \rho \xmat{W})^{-1} \xmat{X}_t \xvec{\beta} \, , \\
	\xvec{\varepsilon}_t = (\varepsilon_t(\xvec{s}_1), \ldots, \varepsilon_t(\xvec{s}_n))' & \sim  & N_n(\xvec{0}, (\xmat{I} - \rho \xmat{W})^{-1} \xmat{\Sigma}_{\varepsilon} (\xmat{I} - \rho \xmat{W}')^{-1} ) \, .
\end{eqnarray*}
Alternatively, the spatial interactions can be modelled in the random effects term in the same manner, while the errors remain independent across space. Generally, there is a relation between geostatistical and spatial autoregression models, and both approaches are equivalent under certain conditions \citep[][Theorem 1]{ver2018relationship}. 

The precision matrix of such spatial autoregressive models is given by 
\begin{equation}
	(\xmat{I} - \rho \xmat{W})' \xmat{\Sigma}_{\varepsilon}^{-1} (\xmat{I} - \rho \xmat{W}) 
\end{equation}
showing the relation to the above-mentioned Cholesky decomposition of geostatistical models. Thereby, the spatial weight matrix $\xmat{W}$ implies a certain (geographical) structure of the Cholesky factors. In this general framework, \cite{zhu2009estimating} proposed a LASSO procedure to estimate the precision matrix, exploiting the fact that geographically distant observations are likely to be conditionally independent (i.e., the precision matrix is a sparse matrix). In this way, the zero entries can be automatically identified.

Instead of the linear relation $\rho \xmat{W}$, the spatial interactions can be modelled using a series of different weight structures, e.g., $\sum_{i = 1}^{k} \rho_i \xmat{W}_i$ with $k$ different weight matrices $\xmat{W}_i$. For instance, each weight matrix could only contain the weights for certain directions (northward, north-eastward, eastward dependence, etc.) to reveal directional processes \citep{merk2021directional}. Moreover, penalised estimation procedures can be used to select the true weight matrix $\xmat{W}$ from a series of alternative weights $\xmat{W}_1, \ldots, \xmat{W}_k$, as for the boosting procedure proposed by \cite{kostov2010model,kostov2013spatial} or the LASSO least-squares procedure proposed by \cite{lam2020estimation}. \cite{ReyesEtAl2012} applied a spatiotemporal LASSO procedure to select weight matrices from a set of candidates with increasing spatial lag order, simultaneously with the temporal lags and spatiotemporal weight matrices. In other words, they constructed a 2-dimensional grid of the temporal and spatial lag orders and selected the relevant spatiotemporal interactions. In the context of traffic analysis, \cite{haworth2014graphical} applied a graphical LASSO procedure to select local neighbourhood structures analogously.  Together with the autoregressive coefficients, \cite{ReyesEtAl2012} also penalised the regressive parameters using a second penalty term. Similarly, \cite{liu2018penalized} suggested a LASSO procedure for selecting the regressors in the mean equation while allowing for spatial autoregressive structure in the model. Since ordinary least-squares procedures are inconsistent in the presence of spatial autoregressive dependence, they proposed a penalised quasi-maximum likelihood approach incorporating a LASSO penalty with a zero shrinkage target for the regression coefficients.

In addition to these approaches, there are several attempts to fully estimate the spatial weight matrix $\xmat{W}$ using regularised procedures. Due to complex interactions and high flexibility, the main issue is to uniquely identify each weight (\citealt{manski1993identification}, and \citealt{gibbons2012mostly} for a critical review of spatial econometric procedures). That is, if one weight between region A and B, say $w_{ab}$, is misspecified, this can be compensated via further linkages through other locations, e.g., via $w_{ac}$ and $w_{cb}$, and still lead to the same spatial covariance matrix. The same applies to the distinction between directed links between A and B and vice versa (i.e., $w_{ab}$ and $w_{ba}$). To the best of our knowledge, \cite{zhu2009estimating} and \cite{bhattacharjee2013estimation} first introduced the idea of estimating the full matrix $\xmat{W}$, where they implied further structural constraints for identification. To be precise, they assumed a triangular weight matrix or symmetric dependence structure. Another structural constraint, namely a block-diagonal structure, was considered in \cite{lam2016detection}. Further, \cite{ahrens2015two} proposed a two-step LASSO procedure to estimate the spatial weight matrix in spatial autoregressive models. For spatial lag models, a regularised estimation procedure was introduced by \cite{lam2020estimation}. Under the assumption of locally constrained spatial dependence, \cite{merk2022estimation} suggested an adaptive LASSO procedure based on cross-sectional resampling, which makes the estimation scalable for large datasets. For spatiotemporal data with unknown structural breaks in the mean, a constraint two-step LASSO procedure was introduced by \cite{otto2022estimation}. Their method could estimate the spatial weight matrix together with all structural breaks and the positions of the change points.





\section{Cross validation under spatiotemporal dependence}\label{sec:cv_strategies}

All regularised estimation procedures require the choice of the degree of regularisation via a so-called penalty parameter, often denoted by $\lambda$. If $\lambda = 0$, the models coincide with their unpenalised version, whereas the degree of penalisation, and thus, the shrinkage of the parameters towards the shrinkage target, increases with an increasing value of $\lambda$.

When a strictly positive $\lambda$ is used, the estimated coefficients $\hat{\beta}_\lambda$ are biased but are more efficient, i.e., their variability is smaller. This is known as the bias-variance trade-off \citep[see, for instance, Section 2.9 of][]{ESL}. Additionally, opting for a $\lambda$ value that is too small can result in overfitting, while selecting a value that is too large can lead to underfitting \citep[see, e.g.,][]{BoonstraEtAl2015}. The optimal penalty parameter is usually selected based on the model's in-sample or out-of-sample predictive performance. 
Other ways to select the penalty parameter are likelihood-based methods, where $\lambda$ is interpreted as a variance component and the likelihood is maximised with respect to the couple $(\sigma_\varepsilon^2, \lambda)'$. Similarly, the penalty parameter can be simultaneously estimated with all parameters in a fully Bayesian approach, where typically vague half-Cauchy prior distribution is assumed for $\lambda$ \citep{vanErpEtAl2019}. Alternatively, \cite{otto2022estimation} proposed to select the penalty parameter based on the distance between the sample and model spatial autocorrelation. Below, we will focus on the predominantly applied goodness-of-fit (GoF) criteria.

GoF criteria aim to maximise the model fit by tuning the penalty parameter $\lambda$. The model fit may be assessed either in-sample or, better, out-of-sample, typically using cross-validation in terms of the distance between the observed and predicted values. The distance is typically evaluated by Root-Mean-Squared-Error (RMSE), Mean Absolute Error (MAE), or based on information criteria, such as Akaike's Information Criterion \citep[][AIC]{Akaike1973} and the Bayesian Information Criterion \citep[][BIC]{Schwarz1978}. Most models have no closed-form solutions to determine the optimal value of $\lambda$, necessitating grid-search algorithms. \cite{ArlotCelisse2010} offers a thorough review of the significance of cross-validation (CV) in regression model selection while providing guidelines to choose the suitable cross-validation procedure according to the specificities of the data, including some brief remarks on dependent data and the problem at hand (e.g., model identification or model selection).



In general, a valid CV must satisfy three properties \citep{JiangWang2017}:
\begin{enumerate}
	\item Randomness of partition;
	\item Mutual independence of test errors;
	\item Independence between the training and test sets.
\end{enumerate}
Regarding the first point, if a model were trained on peak points of seasonal time series and tested on the valley points, the prediction errors would be overestimated. Thus, while periodic partitioning should be avoided, a random partitioning strategy should always be preferred. For illustration of the other two points, consider overlapping folds or partitions. In such a situation, training and test sets would be mutually correlated, as well as the resulting training and test errors. As a direct consequence, the sample variance estimator would remarkably underestimate the actual variance of test errors. Independence among the sets can be assured by leaving a certain \textit{distance} between training and test samples. Thus, for a given test set, all other correlated samples have to be removed from the training set to avoid overfitting 
It means, for example, that when considering spatial or temporal data, the nearby measurements (locations in space or time points) should be removed from the test set when validating a point.
%

Depending on the data structure (e.g., cross-sectional, clustered, spatial, temporal data), as well as on modelling purposes and data-specific features (e.g., seasonality or non-Gaussianity), the cross-validation schemes to be adopted and may vary significantly \citep{ArlotCelisse2010,HewamalageEtAl2023}.
In the case of independent data, classical CV schemes, such as random $k$-fold, stratified $k$-fold \citep{ZengMartinez2000,LudwigEtAl2016} or Generalized CV \citep{BoonstraEtAl2015} can be used. For instance, considering ridge regularisation, the GCV estimator can be used to efficiently estimate the penalty parameter $\lambda$ even when the number of observations (thus, the degrees of freedom) is small or the number of parameters to be estimated exceeds the number of observations \citep{GolubEtAl1979}.
For spatially and temporally dependent data, standard random $k$-fold cross-validation procedures should not be applied. As shown by \cite{SchratzEtAl2019}, while performances are overestimated when no spatial information is included in the CV step, hyperparameter tuning of machine learning models appears to be less sensitive to the spatial structure, leading to similar results of non-spatial and spatial CV schemes.

The problems arising from the occurrence of spatiotemporal autocorrelation in model assessment are manifold. First, using random sampling in cross-validation results in test observations being collected from areas that are spatially close to the training observations \citep{SchratzEtAl2019}. Consequently, the evaluation of prediction performance tends to be overly optimistic because the training and test datasets become correlated, largely due to the neglected underlying correlation structure, whether across space or time \citep{Brenning2012,MeyerEtAl2019,MeyerPebesma2021,PlotonEtAl2020,LezamaEtAl2021}. That is, if one leaves out a set of observations and estimates a spatiotemporal model with the remaining observations, information from the observations used for model estimation is used to predict the left-out observations (via the spatiotemporal interactions). Second, when the spatiotemporal structure is neglected, the estimated residuals will not be mutually independent, which is a critical assumption in many statistical models. As a result, it is often advisable to exclude complete data blocks across time and/or space \citep{RobertsEtAl2017}.


\subsection{Cross-validation in time}

\begin{figure}
	\centering
	\includegraphics[width=12cm ,height=10cm]{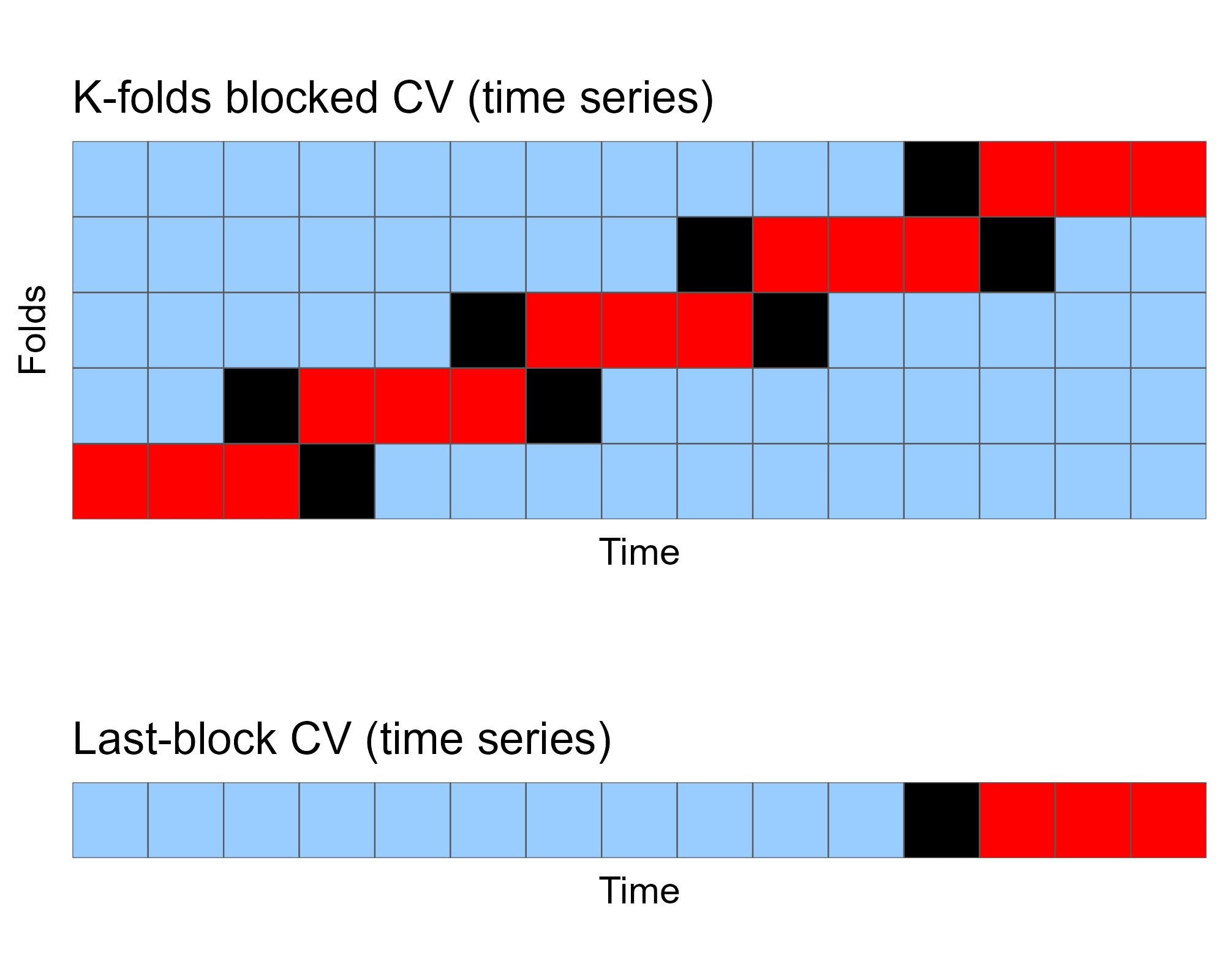}
	\caption{Temporal CV schemes. For a given iteration fold, blue blocks represent time points in the training set, and red blocks represent time points in the test set. Black blocks are the buffer points omitted from both the training and test sets.}
 \label{fig_TSCV}
\end{figure}

Cross-validation is usually applied for stationary time series by segmenting the time series into contiguous subsets to preserve the temporal correlation structure characterizing the observations. Two main strategies are commonly used: (1) $k$-fold blocked subsets CV, where each subset is treated as the test set in turn, and the rest are used for training and forecasting \citep{BergmeirBenitez2012,BergmeirEtAl2014,BergmeirEtAl2018}; and (2) last-block CV or forward validation CV, where only the final block is the test set, and previous blocks are used for training \citep{Hjorth1982}. The key distinction between the two approaches lies in preserving temporal order. Indeed, while the former preserves the natural order of temporal observations by forecasting future values using past \blue{observations} (recall that the model is never tested on past data relative to the training data), the latter uses both past and future values to predict the current test set. In both cases, an $hv$-block method can be used to maintain independence between training and test sets, excluding a window of $h$ observations before and after the test set \citep{Racine2000}. Moreover, both strategies allow choosing among several combinations of forecast horizons and updating schemes for the two samples \citep[e.g., fixed-origin, rolling-origin and recalibration]{Tashman2000} to compute forecasting accuracy metrics \citep{BergmeirBenitez2012}. A graphical synthesis of the two strategies is reported in Figure \ref{fig_TSCV}. Further CV strategies for temporal data can be found in \cite{JiangWang2017} (Markov-CV and partitioned-CV), \cite{CerqueiraEtAl2020} (prequential approach), and \cite{CerqueiraEtAl2017} (Monte Carlo replications last-block approach).

Moreover, time series data are often characterised by non-stationarity of several typologies \citep{HewamalageEtAl2023}. Sources of non-stationarity are seasonality, trends (both deterministic and stochastic), structural breaks or heteroskedasticity. Also, data can be far from Gaussianity due to heavy tails or outliers. In the case of non-stationary time series, the previous CV schemes can be strongly misleading, as the unknown future may differ from the training sample, the test sample, or both. Potential meaningful data splitting strategies include the use of a weighted overlapping approach in which the whole series is used in training and testing steps \citep{BergmeirBenitez2012} or the use of out-of-sample repeated holdout procedures applied in multiple testing periods \citep{CerqueiraEtAl2020}. For an extensive discussion of the role of partitioning schemes for temporal data in a forecasting context, see the recent paper by \cite{HewamalageEtAl2023} in which detailed guidelines on the most appropriate strategy to implement based on the problem to be addressed and the relevant characteristics of the data at hand are provided.

\subsection{Cross-validation in space}
While spatial data share similar challenges related to autocorrelation with time series, the former involves at least two-dimensional coordinates, whereas time is one-dimensional \citep{RobertsEtAl2017}. In addition, \cite{SchratzEtAl2019} demonstrated that hyperparameter tuning in machine learning approaches is less sensitive to spatial structure, but they recommend spatial cross-validation to ensure unbiased predictive performance. Moreover, \cite{MeyerPebesma2021} introduced the ``area of applicability'' (AOA) concept, defining the model's valid geographical area based on training data. That is, the maximum distance (without outliers further apart than 1.5 interquartile ranges) in the covariate space of the training data defines the AOA in the prediction space. All predictions which are further apart than the maximum distance are marked as outside the AOA. The AOA also accounts for the geographical distance if geographical coordinates are included in the covariates. In this way, the idea prevents predicting new geographic spaces with conditions that are very different from the training data, where the models can dramatically fail \citep{MeyerPebesma2022}.

\begin{figure}
	\centering
	\includegraphics[width=16cm ,height=10cm]{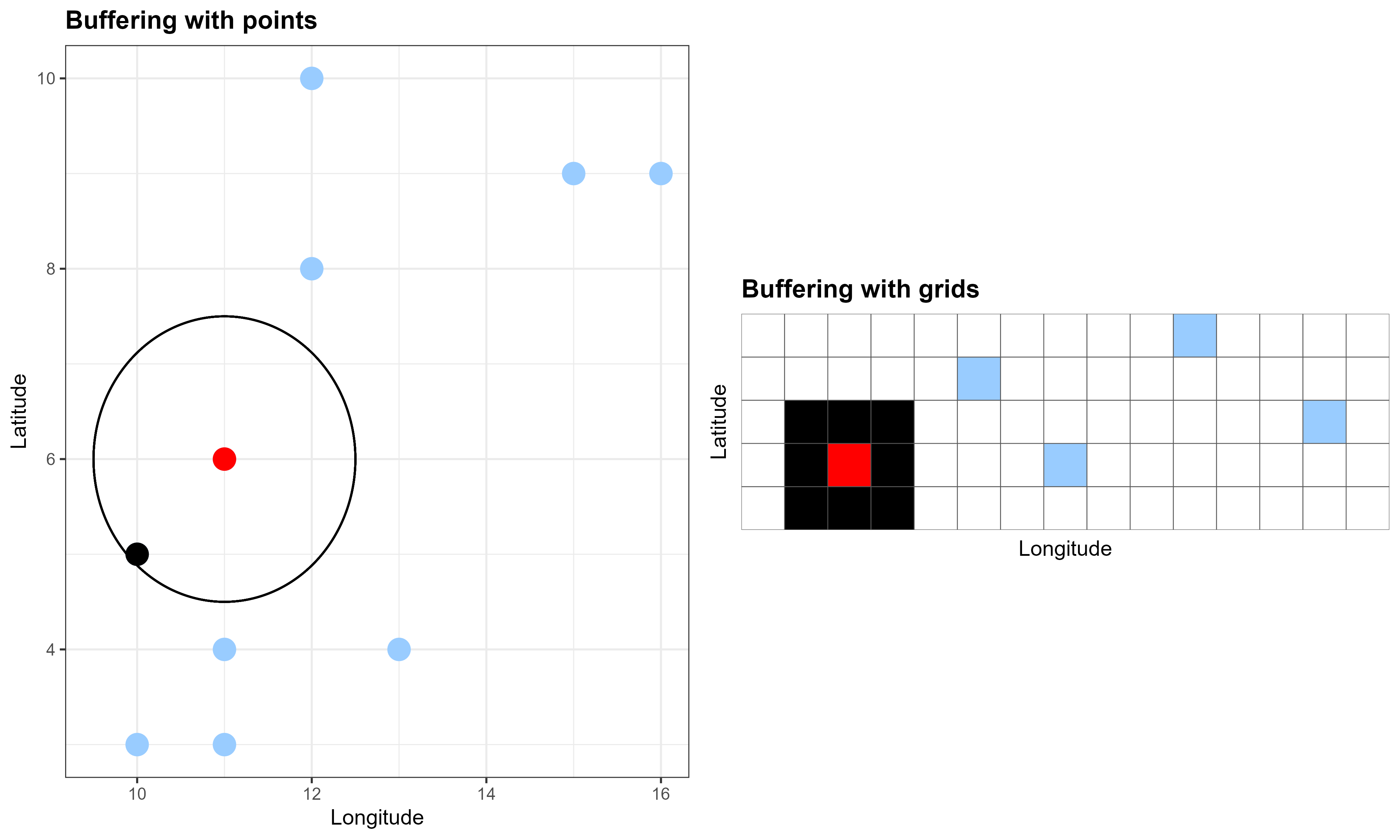}
	\caption{Spatial buffering for point data (left panel) and for gridded data (right panel). The red points represent the test locations, while the blue points represent training locations. Black points represent locations within the buffer of the test set, which are excluded from both the training and test.}
	\label{fig:Buffer}
\end{figure}

To address the above issues, spatial CV methods and spatial variable selection techniques, like recursive and forward spatial feature selection, can be employed  \citep{MeyerEtAl2018,MeyerEtAl2019}. The time-series block CV structure can be easily extended to spatial data by partitioning observations into spatial blocks. These blocks can be created either by dividing the entire space into cells for gridded data or by establishing spatial buffers between the training and test data \citep{RobertsEtAl2017}.
In Figure \ref{fig:Buffer}, we represent examples of spatial buffering both using grids and point data. In the latter, when the CV is performed by iteratively eliminating one location (and the neighbours within the buffer) at a time, we refer to spatial leave-one-location-out (SLOO) CV \citep{GaschEtAl2015,MeyerEtAl2018,MeyerEtAl2019}. As per \cite{LeRestEtAl2014}, SLOO yields a criterion similar to the AIC but without accounting for spatial autocorrelation, meaning it produces the same output as AIC-based model selection in this context. However, when spatially correlated variables are in the model, AIC may not select the right covariates, whereas SLOO performs better. Additionally, spatial blocking can be applied to point patterns by assigning each location to its corresponding training polygon.

When multiple locations are used to build the test set, we refer to spatial $k$-fold CV \citep{PohjankukkaEtAl2017}. To maintain independence between training and test sets, blocks bordering the test set (for spatial blocking) or locations within the buffer (for point data) are excluded from the training set \citep[b-LLO proposed by]{MilaEtAl2022}. The most relevant problem in this situation is to determine what is the optimal buffer length that guarantees independence \citep{TrachselTelford2016}.


Possible proposals include (1) fitting a \textit{prior} variogram to the raw data and using the resulting distance as block length \citep{BioEtAl2002}; (2) estimating the autocorrelation range fitting a variogram \citep{Brenning2005,RobertsEtAl2017,MilaEtAl2022} or a circular variogram \citep{TelfordBirks2009} to the model residuals; (3) implementing a spatial independence test to find the minimum distance such that taken one point in the test set and one in the training set they are uncorrelated \citep{TelfordBirks2005}. However, as pointed out by \cite{Brenning2022}, residuals-based solutions for estimating the autocorrelation range are model-dependent. Thus, the necessity arises for model-agnostic validation tools to assess how predictive performance degrades with increasing prediction distances. The author introduces spatial prediction error profiles (SPEPs), which link the median prediction distances to the spatial prediction errors on the test set. This method can be used to understand (1) how the CV performance of the model decays for increasing distances from the training set and (2) how competing models perform in predicting values at large and small distances.

Alternative approaches to spatial blocking have been proposed in recent geostatistics literature. For instance, geographical partitioning provides a valid alternative via $k$-means algorithm \citep{Brenning2012}. Provided a fixed number of partitions $k$, the clustering algorithm partitions the spatial locations into non-overlapping clusters based on geodesic distance. Unfortunately, the simple $k$-means algorithm does not permit controlling the number of points in each partition, leading to potential heterogeneous partitions. To overcome this problem, \cite{DSTEMv2} proposed a heuristically modified k-means algorithm favouring partitions with similar elements. One may also consider the inverse sampling-intensity weighting system proposed by \cite{deBruinEtAl2022}, in which one can assign more weight to observations in sparsely sampled areas and less weight to observations in densely sampled areas to correct for estimation bias. Even if not explicitly built for treating spatial and temporal CV tasks, potential extensions of the above-cited clustering approaches include the hierarchical spatiotemporal clustering with spatial constraints \citep{ChaventEtAl2018} and spatially-clustered regression \citep{SugasawaMurakami2021}.

\subsection{Cross-validation in both space and time}
When spatiotemporal data are considered, block-based CV schemes are obtained as a combination of the previously cited spatial and temporal blocking strategies. Following \cite{MeyerEtAl2018}, we refer to time-block partitioning as Leave-Time-Out (LTO), point-in-space partitioning as Leave-Location-Out (LLO), and space-time partitioning as Leave-Location-and-Time-Out (LLTO). The three strategies are also called \textit{target-oriented} to contrast the classical \textit{random} approach. Specifically, the LTO partitions the spatiotemporal observations along the time axis into blocks of time series common to all locations (i.e., it performs training by eliminating all time instants assigned to the test block for all spatial locations and iterates over the time blocks). LLO does the same but iterates with respect to the spatial blocks and considers the whole time series of each location. LLTO iterates over both temporal and spatial blocks (i.e., at a given iteration, the algorithm eliminates all observations of a certain spatial block and, for all other blocks, eliminates a common temporal block). Note that LLO coincides with the spatial blocking presented above, whereas LTO coincides with the temporal $k$-fold block strategy.

\begin{figure}[htbp]
	\centering
	\includegraphics[width=18cm ,height=12cm]{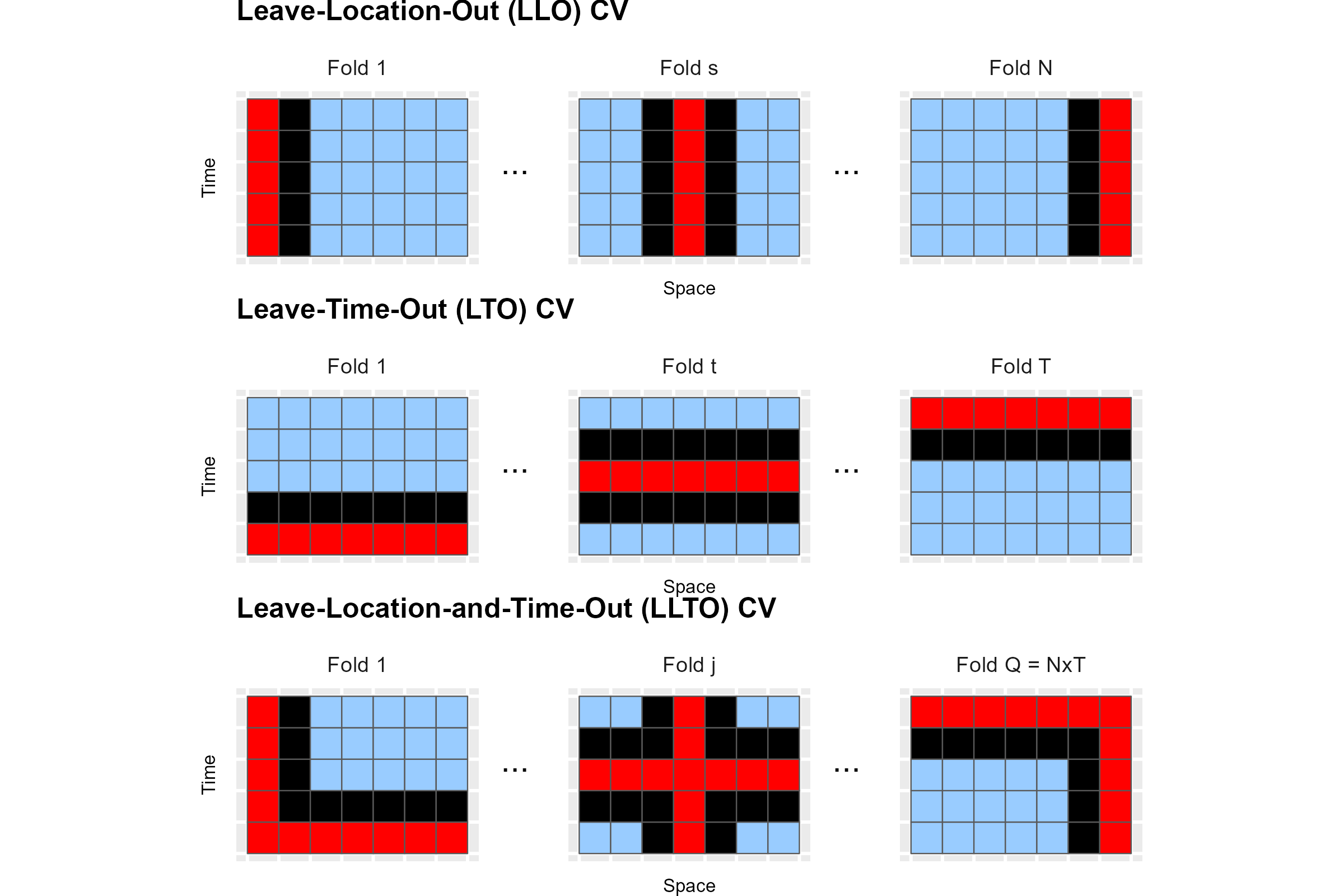}
	\caption{Schematic matrix representation of the spatiotemporal target-oriented CV schemes. In the first row, the LLT algorithm iterates over the columns; in the second row, the LTO algorithm iterates over rows; in the third row, the LLTO algorithm iterates over spatiotemporal cells. Blue cells represent values used for training the model, while red cells represent points in the test set. Black blocks are observations lying within the spatiotemporal buffer that separate the test set from the training set.}
	\label{fig:STCV_synoptic}
\end{figure}

\begin{figure}[htbp]
	\centering
	\includegraphics[width=14cm ,height=10cm]{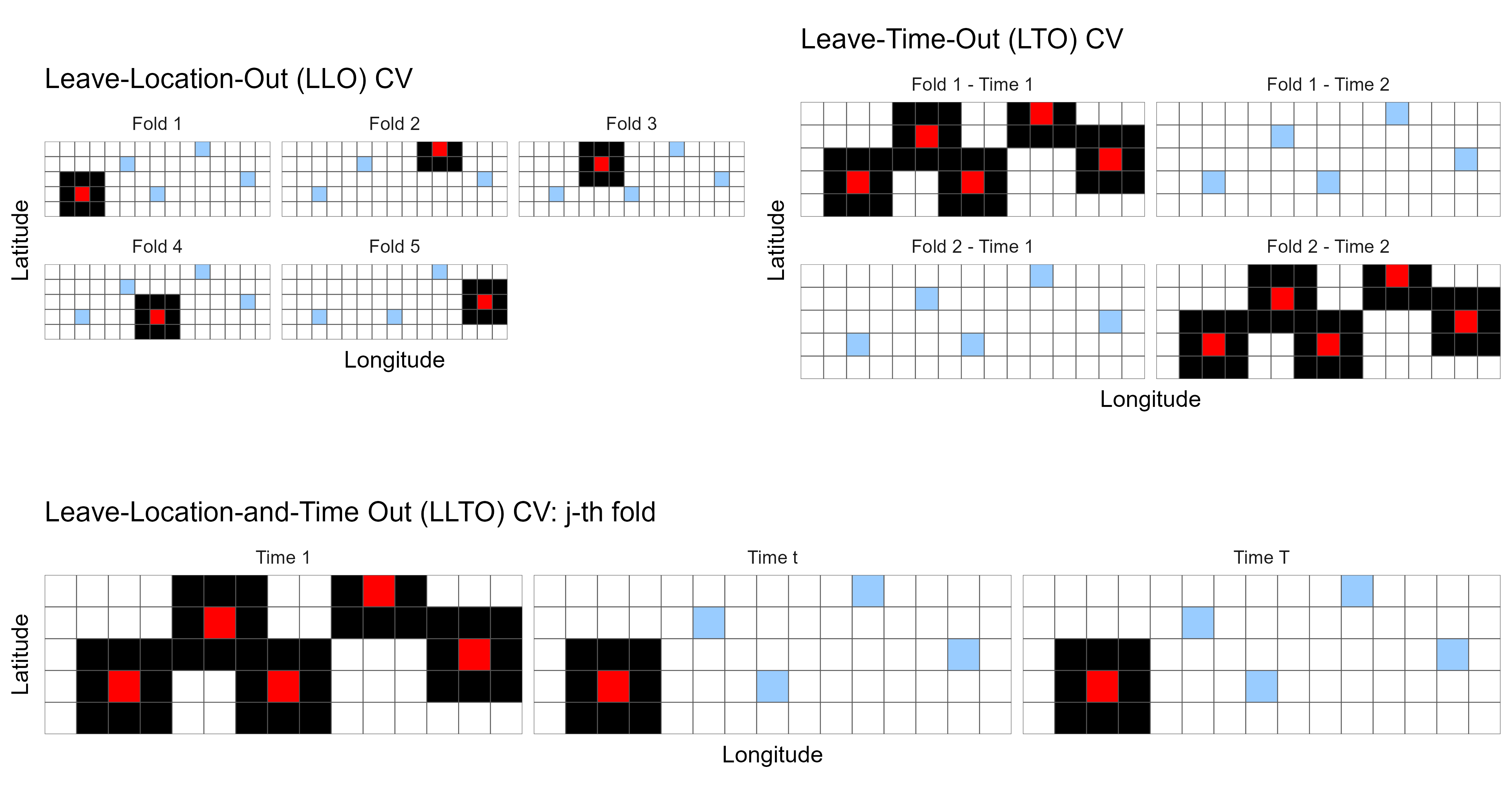}
	\caption{Schematic spatial representation of the spatiotemporal CV schemes. In LTO, the algorithm iterates over rows; in LLO, it iterates over columns; in LLTO, it iterates over cells. Blue cells represent values used for training the model, while red cells represent points in the test set. Black blocks are observations lying within the spatiotemporal buffer and separating the test set from the training set.}
	\label{fig:STCV}
\end{figure}

In Figure \ref{fig:STCV_synoptic}, we show a schematic example of the three above-mentioned target-oriented CV schemes for spatiotemporal data. Time stamps are reported on the rows for each panel, while spatial locations are on the columns. Regarding the latter, assume that each column represents a pair of longitude and latitude values and that stations are ordered according to some distance criterion. Cells marked in red represent values used in the test set, blue cells are the values used to train the model, and black cells are the buffering values used to separate training and test sets. According to the chosen scheme, the algorithm iterates over the rows (LTO), over the columns (LLO) or over the cells (LLTO). A geographical representation of the three algorithms is depicted in Figure \ref{fig:STCV}, in which longitude and latitude are on the x-axis and y-axis, respectively, while each time stamp defines a specific panel. 

A further extension of the LLO approach is the Nearest Neighbour Distance Matching (NNDM) LOO CV introduced by \cite{MilaEtAl2022}. This variant compares the nearest neighbour distance distribution function between the test and training data in the CV process to the nearest neighbour distance distribution function between the target prediction and training points. In practice, this is an alternative method to b-LLO in which the neighbours to be excluded are defined not by distance from the point but by the mismatch between the two Nearest Neighbour Distance distributions. However, as the NNDM algorithm uses an LLO strategy, it is computationally intensive and cannot be used with large datasets. To overcome this issue, \cite{LinnenbrinkEtAl2023} suggested a $k$-fold variant called the kNNDM algorithm.

\subsection{Software for spatiotemporal cross-validation}

Spatial and spatiotemporal CV methods are implemented in many statistical software packages, which we briefly review in this section. Here, we focus on the following main \texttt{R} packages \citep{CRAN2023}: 
\begin{enumerate}
    \item \texttt{sperrorest} \citep{Brenning2012} implements distance-based K-means spatial partitioning,
    \item \texttt{blockCV} \citep{ValaviEtAl2019} implements block partitions and buffering for spatial data, as well as providing geostatistical tools for measuring spatial autocorrelation ranges in candidate covariates for model training and simplifying the choice of block and buffer sizes. It also offers an interactive tool for visualising spatial blocks as a function of folds and block/buffer sizes,
    \item \texttt{CAST} \citep{MeyerEtAl2022,MeyerEtAl2024} implements several spatiotemporal partitioning strategies including NNDM, kNNDMCV, LLO, LTO and LLTO schemes, as well as it allows performing spatial variable selection to select suitable predictor variables according to their contribution to the spatial model performance,
    \item \texttt{mlr3spatiotempcv} \citep{SchratzEtAl2021} is a package being part of the \texttt{mlr3} ecosystem \citep{LangEtAl2019} which provides a unified implementation of a wide range of statistical learning models with feature and model selection tools and model evaluation capabilities. Specifically, \texttt{mlr3spatiotempcv} implements $k$-fold temporal and spatial blocking partitioning with and without buffering described in \cite{MeyerEtAl2018}, and resumes the other partitioning techniques used in \texttt{sperrorest}, \texttt{blockCV}, \texttt{skmeans} \citep[hierarchical agglomerative clustering algorithms by]{ZhaoKarypis2002}, and \texttt{CAST}.
\end{enumerate}

Moving to the \texttt{Matlab} environment, available software for spatiotemporal data partitioning includes the two-fold spatial CV strategy implemented in the DSTEM package \citep{DSTEMv2} for the so-called hidden dynamic geostatistical model. Furthermore, spatiotemporal application of target-oriented CV schemes can be found in \cite{Comparison2024} and \cite{SERRA2023}.






\section{Summary and conclusion} \label{sec:conclusion}

In the field of spatiotemporal statistics and spatial autoregression, this review has demonstrated the vital role of penalised methods in coping with the growing complexity of modern spatiotemporal data. With the increasing availability of geo-referenced data in various formats and types, the demand for adaptable, interpretable, and efficient modelling approaches becomes increasingly evident. This is where regularisation techniques step in, emerging as versatile tools for model selection, dimensionality reduction, and exploring spatial dependencies for classic statistical models. The advantage of interpretability sets them apart from the often enigmatic nature of deep learning models.

In our review paper, we have presented a landscape of different regularisation methods, from the nuances of shrinkage to the mechanisms of penalisation strategies. We have underlined their practicality and effectiveness in the statistical modelling of geospatial data. We also briefly looked at Bayesian regularised estimation methods. Since the regularised estimation methods require the choice of a penalty term, which is usually done by cross-validation, we have also summarised cross-validation methods that can be used in the case of spatiotemporal dependence.

Despite the already substantial literature, integrating and extending regularisation techniques into geostatistical modelling is a promising approach for the future. One notable avenue is the application of regularisation methods for estimating spatial covariance functions, offering novel insights into spatial relationships. A promising approach is estimating spatial covariance functions using regularised splines, which allow an automated choice of basis functions by regularising the smoothness of the estimated function. However, the difficulty lies in ensuring that the covariance function is valid, i.e. that it generates positive-definite covariance matrices.

There is a need for a more extensive exploration of regularised methods in spatial autoregression, particularly in estimating weighting matrices. This matrix is usually assumed to be known, which is rarely the case in practice, but allows the results to be interpreted in a geographical sense. To enhance the interpretability of these estimated matrices, one avenue to consider is using traditional distance-based weighting matrices as shrinkage targets, thereby enabling a geographical interpretation. Generally, the quest for enhanced computational efficiency in the application of regularisation methods for large-scale spatiotemporal applications remains a pertinent concern in both fields.

In summary, in a world where the diversity and volume of geospatial data continue to increase, this review paper is intended to provide guidance for understanding regularised methods in spatial and spatiotemporal statistics to advocate their use for geospatial analyses.

\end{document}